\documentclass[11pt]{article}
\usepackage{amsfonts, amssymb, amsmath, graphicx, epsfig, lscape, subfigure}
 \pdfoutput=1

\def\inst#1{$^{#1}$}

  {\hspace*{\fill}$\Box$\par\vspace{4mm}}
  {\hspace*{\fill}$\Box$\par\vspace{4mm}}
  {\hspace*{\fill}\par\vspace{4mm}}

\newcommand{\qed}{$\Box$}

\begin {document}

\title{Resilience of Core-Periphery Networks in the Case of Rich-Club}

\author{%
Matteo Cinelli\inst{1} \and Giovanna Ferraro\inst{1} \and Antonio
Iovanella\inst{1}%
}

\date{}

\maketitle

\begin{center}
{\footnotesize 
\inst{1} Department of Enterprise Engineering\ \

University of Rome ``Tor Vergata''\\
Via del Politecnico, 1 - 00133 Rome, Italy.\\
\texttt{matteo.cinelli@uniroma2.it\\
giovanna.ferraro@uniroma2.it\\
antonio.iovanella@uniroma2.it\\}}
\end{center}

\date{}

\begin{abstract}
Core-periphery networks are structures that present a set of central and densely connected nodes, namely the core, and a set of non-central and sparsely connected nodes, namely the periphery. The rich-club refers to a set in which the highest degree nodes show a high density of connections. Thus, a network that displays a rich-club can be interpreted as a core-periphery network in which the core is made up by a number of hubs. In this paper, we test the resilience of networks showing a progressively denser rich-club and we observe how this structure is able to affect the network measures in terms of both cohesion and efficiency in information flow. Additionally, we consider the case in which, instead of making the core denser, we add links to the periphery. These two procedures of core and periphery thickening delineate a decision process in the placement of new links and allow us to conduct a scenario analysis that can be helpful in the comprehension and supervision of complex networks under the resilience perspective. The advantages of the two procedures, as well as their implications, are discussed in relation to both network efficiency and node heterogeneity.

\end{abstract}

%%%%%%%%%%%%%%%%%%%%%%%%%%%%%%%%%%%
\section{Introduction}
\label{Intro}

Defined as a system's ability to adjust its activity to retain its basic functionality when errors, failures and environmental changes occur \cite{cohen2000resilience, gao2016universal}, resilience is a crucial property of many networked systems. It has been rapidly tackled by the scientific literature \cite{albert2000error, crucitti2004error} and, as such, is still considered a topic of great interest \cite{dedomenico2017modeling, gao2016universal}.

Related to concepts such as robustness, redundancy, vulnerability and sustainability \cite{okelly2015networkhub}, resilience is considered fundamental for a number of practical approaches that involve risk assessment in terms of criticalities related to the eventual failure (or removal) of nodes and links and thus by means of overall systemic tolerance.
Indeed, network performances (especially in terms of routing ability and stability) are directly related to their resilience and thus to the capabilities of networks in tolerating loss of important elements such as bridges or hubs.
Mainly because of its tangible implications \cite{modica2015spatial, rubinov2010complex, williamson2003resilient, zhao2011analyzing}, resilience has been investigated across many different network structures (both synthetic and real) and there is now knowledge regarding how specific kinds of networks react to specific kinds of losses \cite{holme2002attack, iyer2013attack}.
%For instance, scale free networks are particularly susceptible to failures regarding hubs while they are heavily resilient to random failures)  
In more detail, since resilience is related to the ability to withstand deliberate attacks and incidents, studies about this topic have tended to consider a large variety of structural failures (both induced by attack or naturally occurring) which involve both specific (i.e. chosen by their properties like the centrality indexes) and random nodes.

Moreover, as resilience is strictly related to the network topology \cite{albert2000error, crucitti2003efficiency}, results of the stress tests 
%from the former perspective (i.e. from the resilience perspective) 
are strongly affected by certain structural measures such as density and the clustering coefficient \cite{FI2017}, as well as by the presence of specific substructures like cliques or dense subgraphs, which are, in general, highly fault-tolerant since the loss of any element has no disruptive effect on the interaction between the others. 

Among those densely tied substructures that seem to be of interest in terms of resilience \cite{gutfraind2010optimizing, yang2015improving}, the rich-club is particularly well known \cite{zhou2004rich}.
The rich-club is a network substructure that is observed when hubs are tightly interconnected. It constitutes the basis for the recognition of the rich-club phenomenon which is, more generally, defined as the tendency of nodes with a high centrality (usually degree) to form highly interconnected communities \cite{colizza2006detecting}. Furthermore, it can be even interpreted as the core of a core-periphery network \cite{ma2015richcores}, i.e. as the core of a network that shows a set of central and densely connected nodes and a set of non-central and sparsely connected nodes. 

The rich-club phenomenon has been observed in many different networks \cite{colizza2006detecting, zhou2004rich} and its importance has been recognized in that it represents an unexpected feature ( i.e. non-replicated by regular models \cite{csermely2013structure, zhou2004rich}) of many real systems, which is shown to have a relevant effect on certain network measures, especially on assortativity and transitivity \cite{xu2010rich}.
Another important aspect of the rich-club is that, while it is possible to evaluate its presence for each value of the node degrees, through a specific coefficient properly normalized over an ensemble of randomized networks \cite{colizza2006detecting, csigi2017geometric, jiang2008statistical, cinelli2017rich, zhou2012random, zhou2007structural}, it is not possible to compute its size a priori~\cite{cinelli2017rich}. 

Thus, it is commonly assumed that the rich-club is made up of a certain low percentage of the highest degree nodes \cite{xu2010rich, zhou2004rich}, whose interconnections are able to strongly affect a number of structural measures.
So, despite the fact that a number of studies have investigated the rich-club phenomenon and aspects of resilience within the context of complex networks (like in the case of the Internet \cite{cohen2000resilience} and, more recently, of the Darknet \cite{dedomenico2017modeling}), to the authors' knowledge these two problems have never been tackled when taking their conjectured mutual effects into consideration.
Indeed, while there have been some statements about the role of the rich-club in terms of its capacity to increase the network stability \cite{dedomenico2017modeling}, to act as a super traffic hub \cite{zhou2004rich} and to indicate resilience to specific kind of attacks \cite{piraveenan2014nodeassortativity}, the literature still lacks of a unique general framework able to explicit the relationship between the rich-club ordering and the resilience of a network.

Under these circumstances, this paper aims to shed some light on the role of the rich-club from a resilience perspective by looking at how the presence and the characteristics of this important substructure are able to affect the network robustness from various points of view.       

For these reasons, we analyze resilience by considering networks in which we manipulate the set of connections among the highest degree nodes by adding and removing links. By adopting this strategy we obtain a set of different networks that share the same topology other than a small subgraph made up of the rich nodes, i.e. we keep the network periphery while altering the network core. The resilience is tested on the resulting networks by means of a number of measures related to both efficiency and cohesion: the diameter, the average path length, the global efficiency and the global clustering coefficient. 
The implications of the rich-club presence in terms of resilience lay the basis for the investigation of a different rationale in the positioning of new links. Therefore, we modify the previous manipulation procedure by testing the case in which the same amount of links (which we would add in order to reach certain rich-club densities) is instead added randomly outside the rich-club. 

More specifically, we implement  two procedures of either core or periphery thickening in order to mimic the decision process of a supra agent that, with a limited amount of resources constituted by the new links, has to engineer the considered system in an efficient manner. The result of this process will be relevant in understanding where to put new connections in existing networks, such as new routes in airport networks or new cables in power grids or the Internet, being consistent with a set of efficiency criteria that are here represented by the network measures used in the evaluation of resilience. Lastly, our results allow room for certain considerations at different levels, which will be useful in better comprehending and supervising networks that display the rich-club structure.   

The paper is organized as follows: Section~\ref{Richres} describes rich-club ordering and network resilience; Section~\ref{Simset} shows the simulation setting; Section~\ref{probset} displays the simulation results and analysis; Section~\ref{conclusion} presents discussions and conclusions.

%%%%%%%%%%%%%%%%%%%%%%%%%%%%%%%%%%%
\section{Rich-Club Ordering and Network Resilience}\label{Richres}
%\subsection{First Subsection}

Rich-club ordering is an important topological property firstly observed in the case of technological networks and, in more detail, in the case of the Internet at Autonomous Systems (AS) level \cite{zhou2004rich}. Recognition of this phenomenon is conducted via a comparison between the number of links among the rich nodes and the number of links they might possibly share. In doing so, it is possible to evaluate the density of the subgraph made up of such nodes. The rich nodes are those that have a degree higher than a certain threshold $k$ and a rich-club occurs when such nodes are more densely interconnected than expected, i.e. they have more interconnections with respect to the average of the interconnections found among the same nodes in an ensemble of rewired networks \cite{colizza2006detecting}.

However, as the threshold value of degree $k$ for which we may observe the rich-club is unknown, the size of the rich-club is therefore assumed, in accordance with the empirical evidence, to be around the $1\%$ of the network nodes \cite{cinelli2017rich, xu2010rich, zhou2004rich}. The empirical evidence of small rich-club size is present in many different domains from technological \cite{zhou2007structural} to social \cite{masuda2006vip} and biological networks where, especially in neuroscience \cite{sporns2004organization, van2012high, van2011rich}, the investigation of the rich-club phenomenon has provided important insights from a brain functionality perspective. 

Thus, while this property has been recognized as relevant, its effect
%, in terms of presence or absence of the subgraph, 
on the network metrics has been mainly tested for cohesion measures such as the clustering coefficient and the degree assortativity, and only marginally for path-based measures that should be, in case of rich-club ordering, more of interest since such measures are associated with information flow.
Indeed, the efficiency of a network is mainly based on path metrics and it has been shown that the rich-club is an emergent property of certain networks \cite{csigi2017geometric} in which hubs need to be interconnected in order to avoid losses, as in the case of electric current in power grid networks \cite{csigi2017geometric, simpson2016voltage}. In this respect, the knowledge and the investigation of the rich-club effect on other measures, closer to the concept of distances among nodes, may be of interest in terms of both static analysis, i.e. in terms of the effect of a progressively denser rich-club on certain measures, and dynamic analysis, i.e. in term of resilience.
Indeed, the investigation of network resilience can be seen as a what-if analysis that considers a large set of network topologies and metrics that derive, through a procedure of nodes and links deletion, from the original one.

Resilience has been traditionally studied in two different cases (or scenarios): error and attack. By error we mean the random removal of elements; by attack we mean a removal process that targets specific or crucial elements. Thus, the error case considers randomness while the attack case is conducted by removing elements with high values of certain centrality measures in two different ways: sequential and simultaneous \cite{iyer2013attack}.
If we consider node removal, in the sequential targeted attack the centrality measures are computed after each node removal and the node with the greater centrality score is eliminated; in the simultaneous targeted attack the centrality measures are computed at the beginning and the order of the nodes to be removed is known before the procedure starts.
In the previous cases and even in the case of error, the basic properties and effect of the removal procedures are well known in the literature for both real and synthetic networks \cite{holme2002attack, iyer2013attack}. For instance, it is known that scale-free networks are particularly resilient in case of error and particularly vulnerable in case of attack due to the variance of their degree distribution, i.e. because of a topology that includes hubs \cite{albert2000error}.
% Additionally these effects are kept in case of simultaneous degree targeted attack but not in the case of sequential degree targeted attack as in this case the removal by betweenness or eigenvector are proved to be more disruptive.
Obviously, many other cases could be mentioned, but none of them would include, to our knowledge, a clear perspective on the role of the rich-club in such networks. Thus, under these circumstances and given the relevance of both network resilience and rich-club ordering from a number of perspectives, it is important to extend the current knowledge as deriving from the literature to the case of networks displaying a rich-club structure.

\section{Simulation Setting}\label{Simset}
%\subsection{First Subsection}
\label{Simset1}

We analyze resilience by considering undirected and unweighted scale-free networks $G$, with $N = 5000$ nodes and mean degree value $\langle k \rangle = 6$. We manipulate the connections among the top 1\% nodes of highest degree by adding/removing links in order to create subgraphs (cores) with various density values. In adopting this strategy we are able to obtain different networks sharing the same topology other than the subgraph made up of the rich nodes. 

\begin{figure}[tbh]
% \begin{minipage}[b]{6.5cm}
 \begin{center}
\begin{tabular}{@{}cccc@{}}% qualunque cosa significhi
% http://tex.stackexchange.com/questions/84889/combining-multiple-eps-files-into-a-single-figure
 \includegraphics[scale = 0.25, trim=3.5cm 1cm 2.5cm 0cm, clip=true]{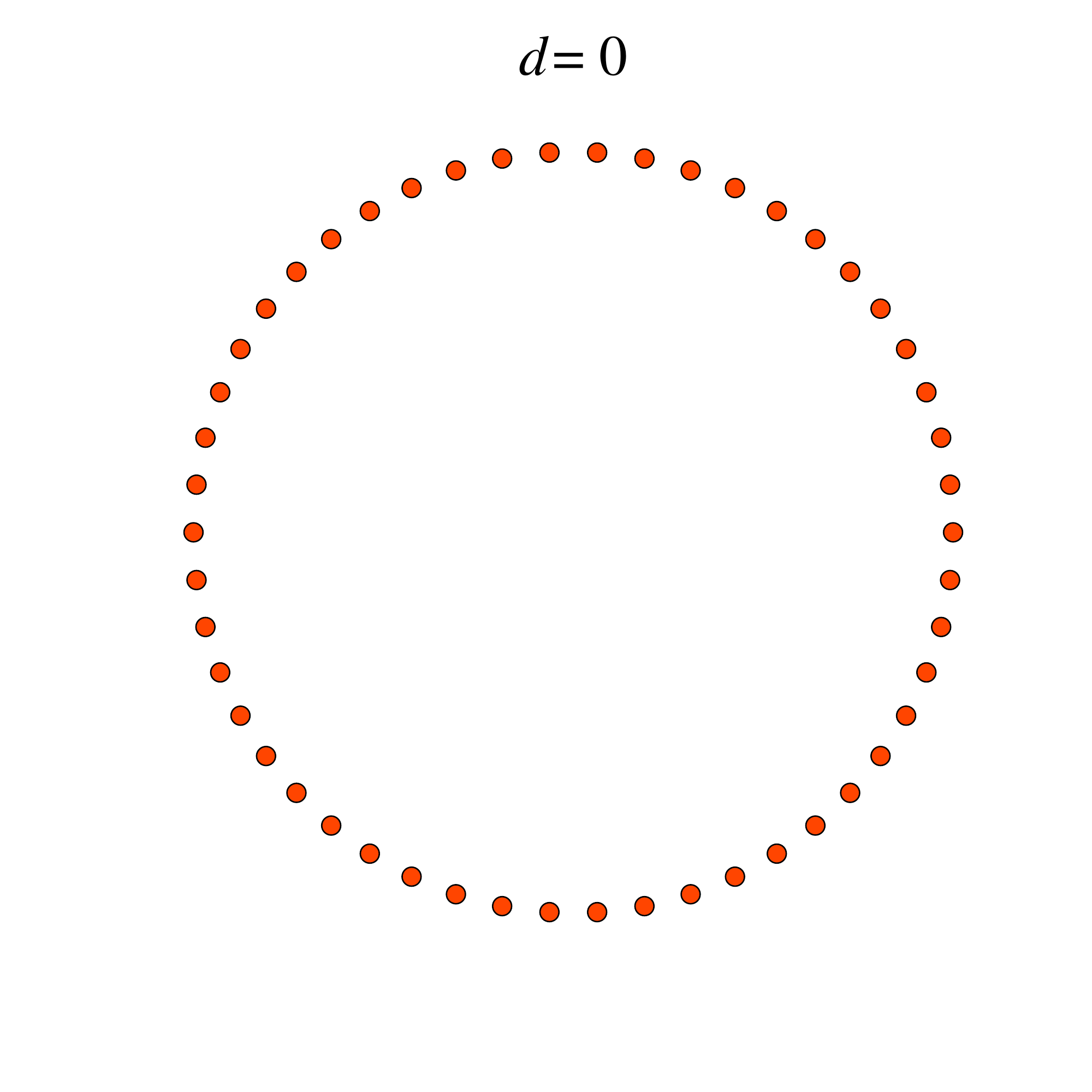} & 
 \includegraphics[scale = 0.25, trim=3.5cm 1cm 2.5cm 0cm, clip=true]{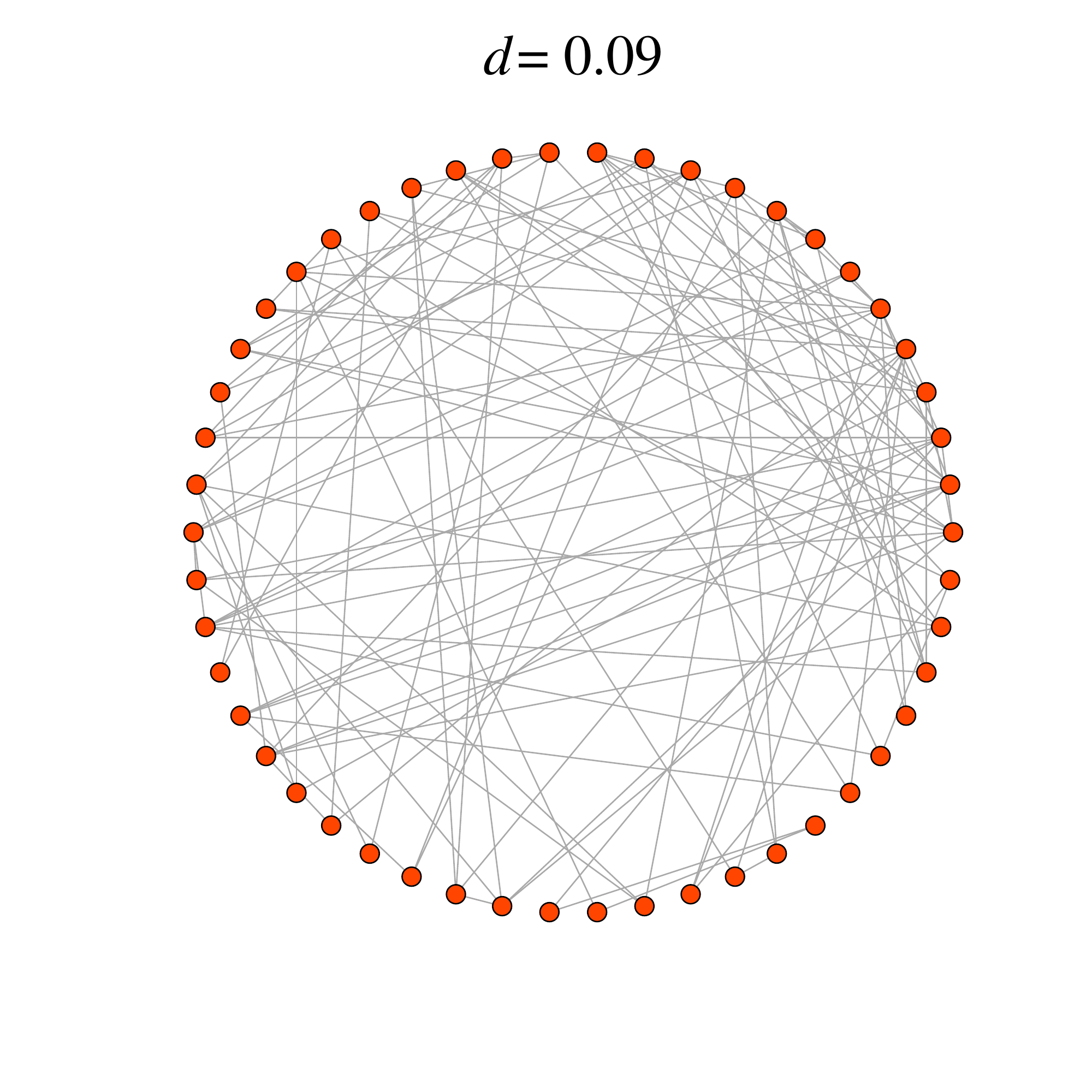} &
 \includegraphics[scale = 0.25, trim=3.5cm 1cm 2.5cm 0cm, clip=true]{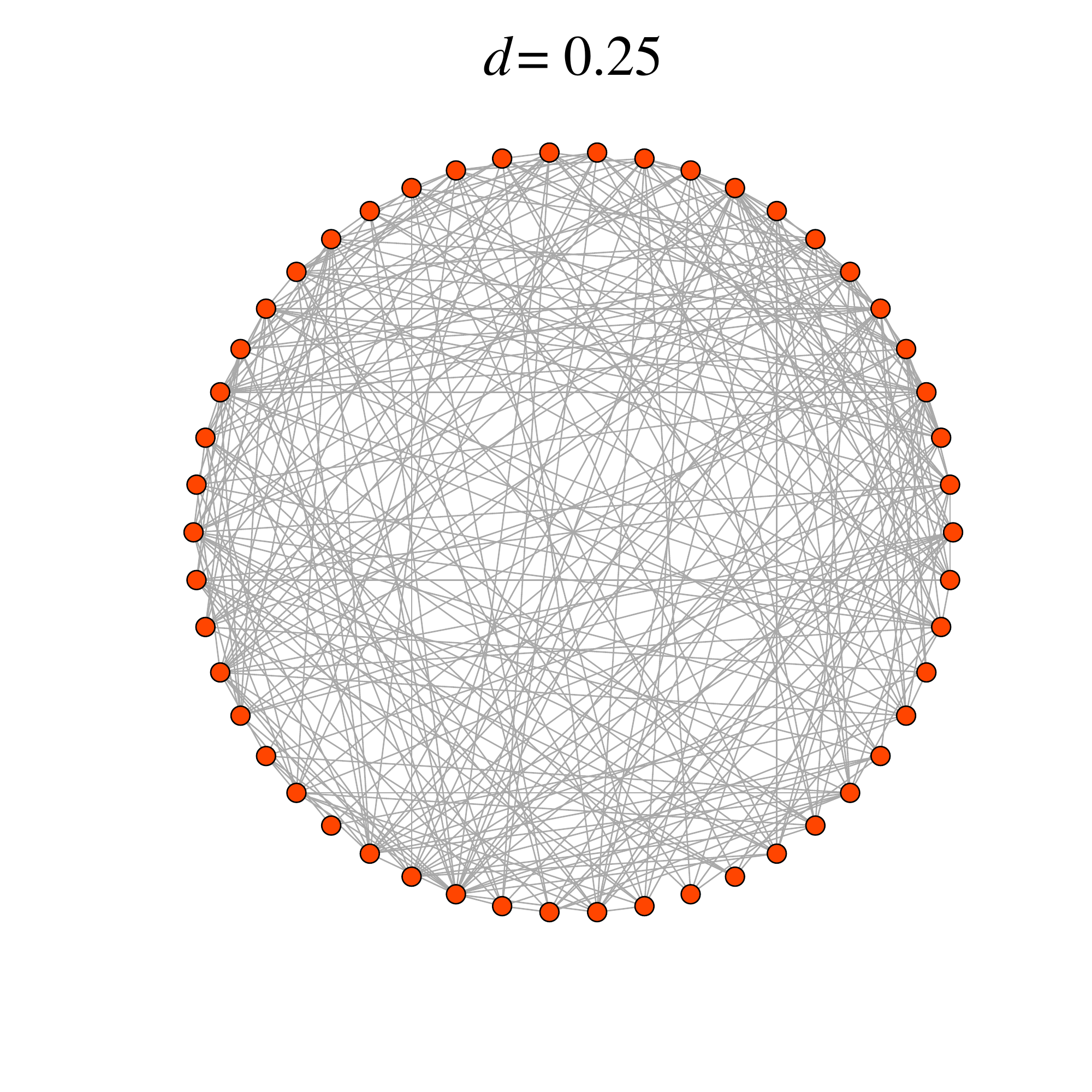} \\ 
 \includegraphics[scale = 0.25, trim=3.5cm 1cm 2.5cm 0cm, clip=true]{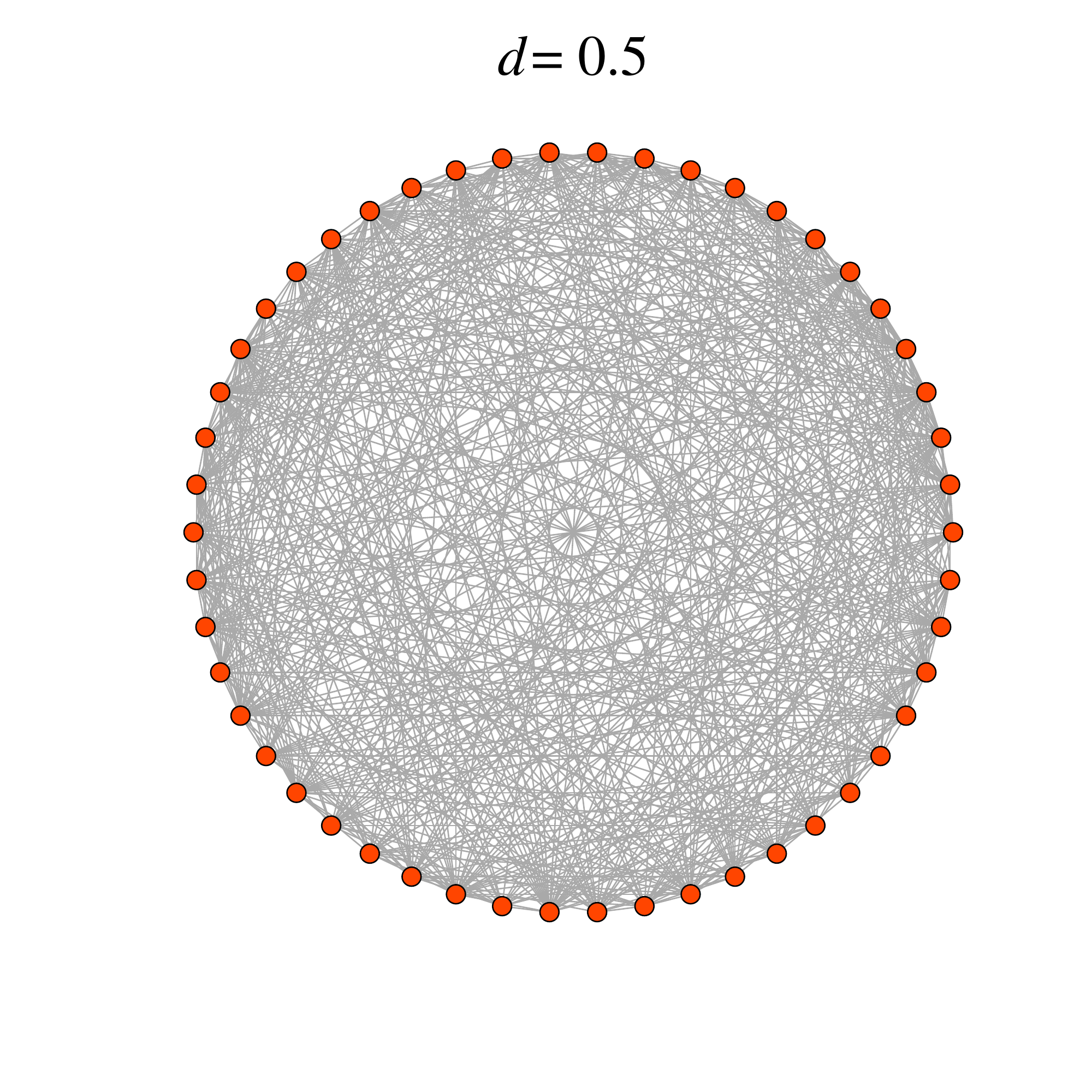} & 
 \includegraphics[scale = 0.25, trim=3.5cm 1cm 2.5cm 0cm, clip=true]{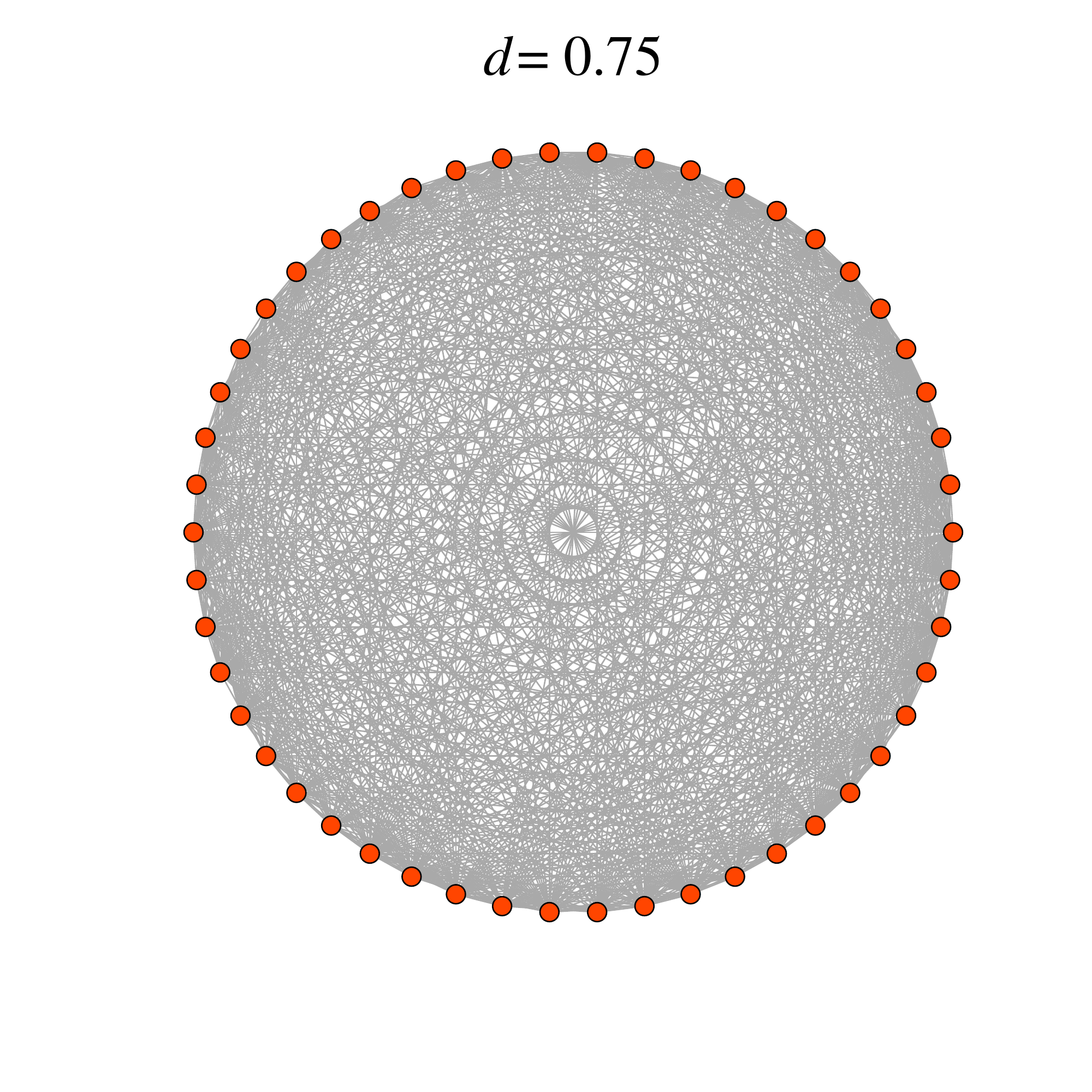} &
 \includegraphics[scale = 0.25, trim=3.5cm 1cm 2.5cm 0cm, clip=true ]{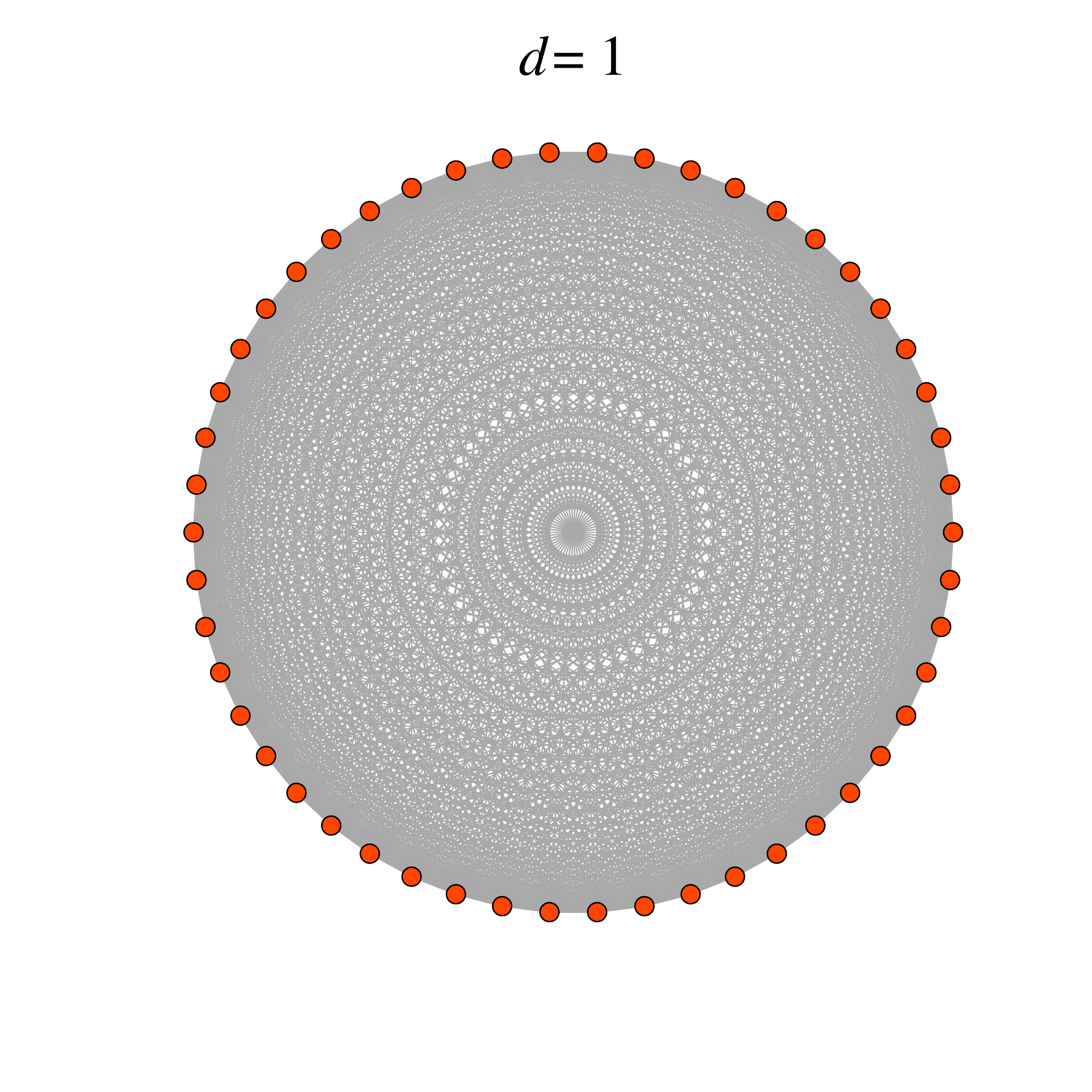} 
\end{tabular}
\end{center}
%\end{minipage}
\caption{Process of link addition/removal of the subgraph made up of the highest degree nodes in order to reach different density values.}
\label{fig_config}
\end{figure}

As  shown in Figure \ref{fig_config}, the obtained densities of the induced subgraphs are $d = \{0, 0.09, 0.25, 0.5, 0.75, 1\}$ where $d = 0.09$ is the density, averaged over ten instances, of the subgraph made up of rich nodes in the original (i.e. non-manipulated) network. This last case represents the default case among the different generated networks. In the six different scenarios we test the robustness of the network to node removal in case of error and in case of simultaneous degree-targeted attack. The choice of this kind of attack (instead of the sequential degree-targeted attack in which the centrality scores are computed at each iteration) is motivated by the fact that our aim is to observe the effect of the rich-club, as realized by our manipulation, on certain measures that characterize the considered network. 
Indeed, with the simultaneous degree-targeted attack we know a priori the nodes that are going to be removed, while in case of sequential degree targeted attack the ensemble of rich nodes may be subjected to variations due to the re-computation of the centrality scores at every iteration.
%are sure that the first x\% of the removed nodes, where x\% is the number of rich nodes divided by the number of network nodes, are those that constitute the rich-club.

After the removal of each node, we compute a number of different metrics that refer to aspects of both information flow and network cohesion. The considered measures are global in the sense that they are computed on the whole network and not on the single node, and they are: the diameter, the average path length, the global network efficiency and the global clustering coefficient (see Table~\ref{tab_measure}). The obtained results are averaged over 10 replicas of the resilience tests and on 10 different networks realized using the same degree sequence (i.e. the same list of node degrees). 

\begin{table}[t]
\begin{footnotesize}
%\begin{tiny}
\begin{center}
\begin{tabular}{|l|p{6cm}|c|}
        \hline
Measure & Definition & Formula\\
\hline
\hline
Diameter ($D$) & The length of the shortest path between the most distanced nodes. & $D = \max\limits_{i,j \in G}d_{ij}$\\
Average path length ($APL$) & The mean of all the shortest paths between all couples of nodes. & $APL = \frac{1}{N(N-1)}\sum\limits_{i, j \in G}d_{ij}$\\
Global network efficiency ($E$) & A measure of how efficiently the network exchanges information. & $E = \frac{1}{N(N-1)}\sum\limits_{i \neq j 
\in G}\frac{1}{d_{ij}}$\\
Global clustering coefficient ($\overline{C}$) & The average of the local clustering coefficients $C_i$ of all individual nodes. & $\overline{C} = \frac{1}{N}\sum\limits_{i \in G} C_i$\\
\hline
\end{tabular}
\end{center}
\caption{Short glossary of metrics computed during simulations (note that $d_{ij}$ is the shortest path between nodes $i$ and $j$ in $G$).}\label{tab_measure}
%\end{tiny}
\end{footnotesize}
\end{table}

For all the considered cases we focus on the initial effect of a denser/sparser rich-club on the measures from above and on its effect throughout the process of node removal. Additionally, we test the case in which the same amount of links that we would add in order to reach certain rich-club densities are instead added randomly outside the rich-club. In other words, by recalling the core/periphery nature of networks that display rich-club ordering, we test two procedures of either core or periphery thickening.
The comparison between the two procedures allows us to perform a scenario analysis and to simulate a decision process of a supra agent that, with a limited amount of resources (the links), has to engineer the considered system (the network) in an efficient (from the point of view of the described measures) manner. % STA PARTE RIBUTTALA NELL'INTRODUZIONE O NELL'ABSTRACT

\begin{figure}[tbh]
\begin{center}
\includegraphics[scale = .4]{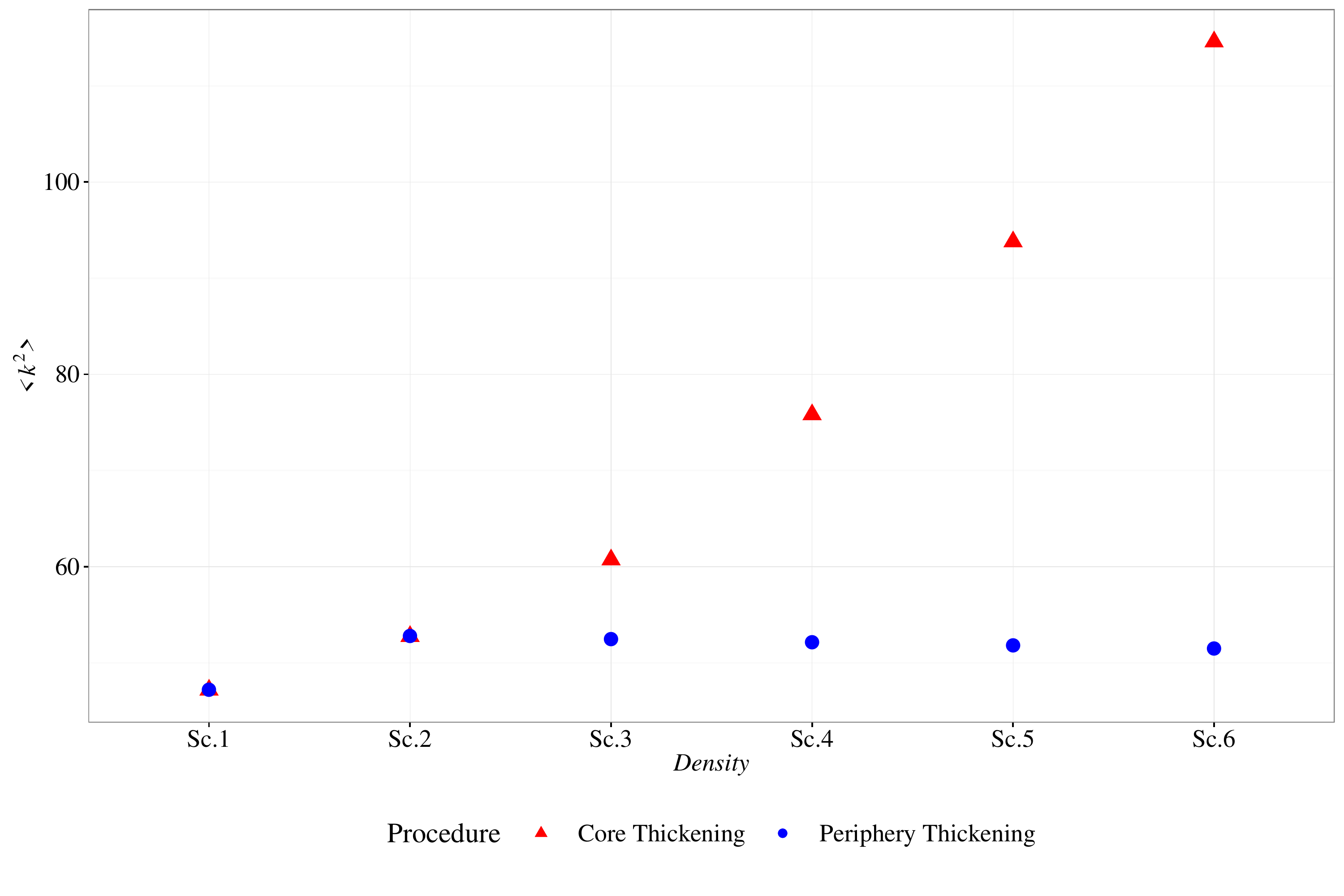}
%trim=3.5cm 24.3cm 3cm 3cm, clip = true
\caption{Variance of the degree $\langle k \rangle ^2$ after the procedure of core and periphery thickening. All results are averaged on 10 instances.}
\label{var}
\end{center}
\end{figure}

It is worth to adding at this point that the two indicated procedures $i)$ to add links within the rich-club, and $ii)$ to add links outside the rich-club, alter the degree distribution (and the degree sequence) of the considered networks.
These alterations depend on many factors, including the number of links to be added and their location, as well the consequent size and density of the rich-club. Placement of the new links has an effect on the different portions of the degree sequence, meaning the two procedures end up turning the network into either a more irregular or regular structure. We illustrate this process of degree sequence modification by plotting the variance $\langle k \rangle ^2$ of the node degrees (see Figure \ref{var}), i.e. the degree-related network heterogeneity \cite{jacob2017measure, snijders1981degreevariance}, in the described cases.    

We summarize the described simulation procedures as shown in Table~\ref{tab_scenario} where column 2 is the required local density for the rich-club subgraph, column 3 is the average values of links to be added in order to obtain such a density, column 4 reports the number of links removed or added randomly in the network core, while column 5 highlights the number of links that are randomly removed or added in the network periphery. Note that links are reported as averages over ten instances, while in the network manipulation each of the ten instances were modified with the proper number of links. Note also that in the second setting the default rich-club is preserved together with its density, since we are adding links in the network periphery.

Data processing, the network analysis and all simulations were conducted using the software {\it R}~\cite{R} with the {\it igraph} package~\cite{csardi2006igraph}.

%\bibitem[\protect\citeauthoryear{R Core Team, 2014}{}]{RCT}
%R Core Team (2014) R: A Language and Environment for Statistical Computing, {\it R Foundation for Statistical Computing}, Vienna, Austria, http://www.R-project.org.
%
%\bibitem[\protect\citeauthoryear{Csardi and Nepusz, 2006}{}]{CN}
%Csardi, G. and Nepusz, T., (2006) `The igraph software package for complex network research', {\it InterJournal Complex System}, vol. 1695, http://igraph.org.

\begin{table}[t]
\begin{footnotesize}
%\begin{tiny}
\begin{center}
\begin{tabular}{|l|c|c|p{3cm}||p{3cm}|}
        \hline
                  & Rich-club density & Links & Core thickening & Periphery thickening\\
\hline
\hline
Scenario 1 & $d_{rc} = 0$      & $\overline{m}_1 = 111$ & Remove $m_1$ links & Remove $m_1$ links \\
Scenario 2 & $d_{rc} = 0.09$ & $\overline{m}_2 = 0$ & Default case & Default case \\
Scenario 3 & $d_{rc} = 0.25$ & $\overline{m}_3 = 194$ & Add $m_3$ links & Add $m_3$ links\\
Scenario 4 & $d_{rc} = 0.50$ & $\overline{m}_4 = 500$ & Add $m_4$ links & Add $m_4$ links\\
Scenario 5 & $d_{rc} = 0.75$ & $\overline{m}_5 = 807$ & Add $m_5$ links & Add $m_5$ links\\
Scenario 6 & $d_{rc} = 1$      & $\overline{m}_6 = 1113$ & Add $m_6$ links & Add $m_6$ links\\
\hline
\end{tabular}
\end{center}
\caption{Simulation scenarios for core and periphery thickening.}\label{tab_scenario}
%\end{tiny}
\end{footnotesize}
\end{table}

\section{Simulation results}\label{probset}
\subsection{Core Thickening}

\begin{figure}[tbh]
\begin{center}
\includegraphics[scale = .5]{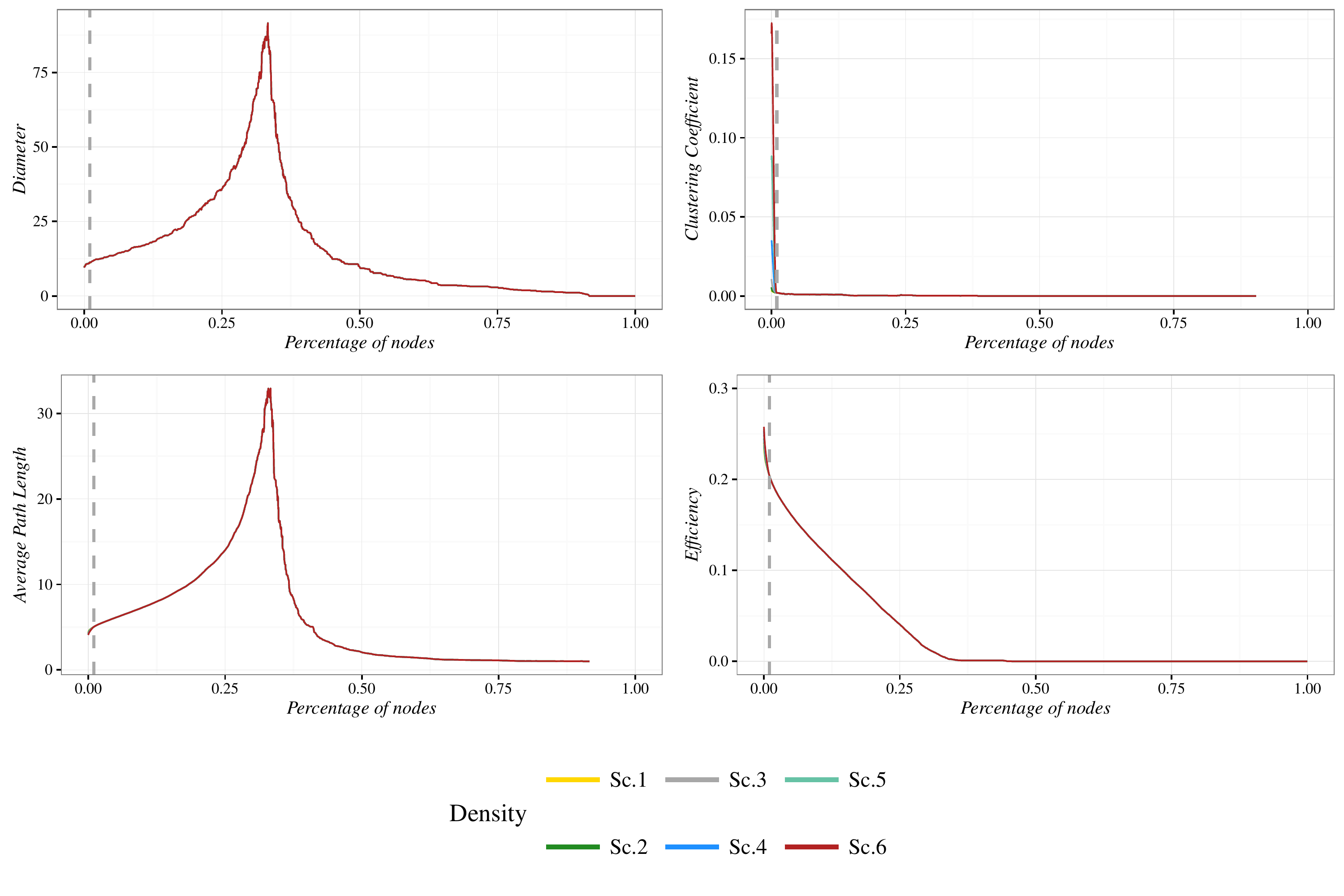}
%trim=3.5cm 24.3cm 3cm 3cm, clip = true
\caption{Resilience for simultaneous attack simulations with progressive manipulation of the number of links in the network core. 
The dashed line is placed in correspondence to the rich-club size.
All results are averaged over 10 instances.}
\label{attack}
\end{center}
\end{figure}

\begin{figure}[tbh]
\begin{center}
\includegraphics[scale = .5]{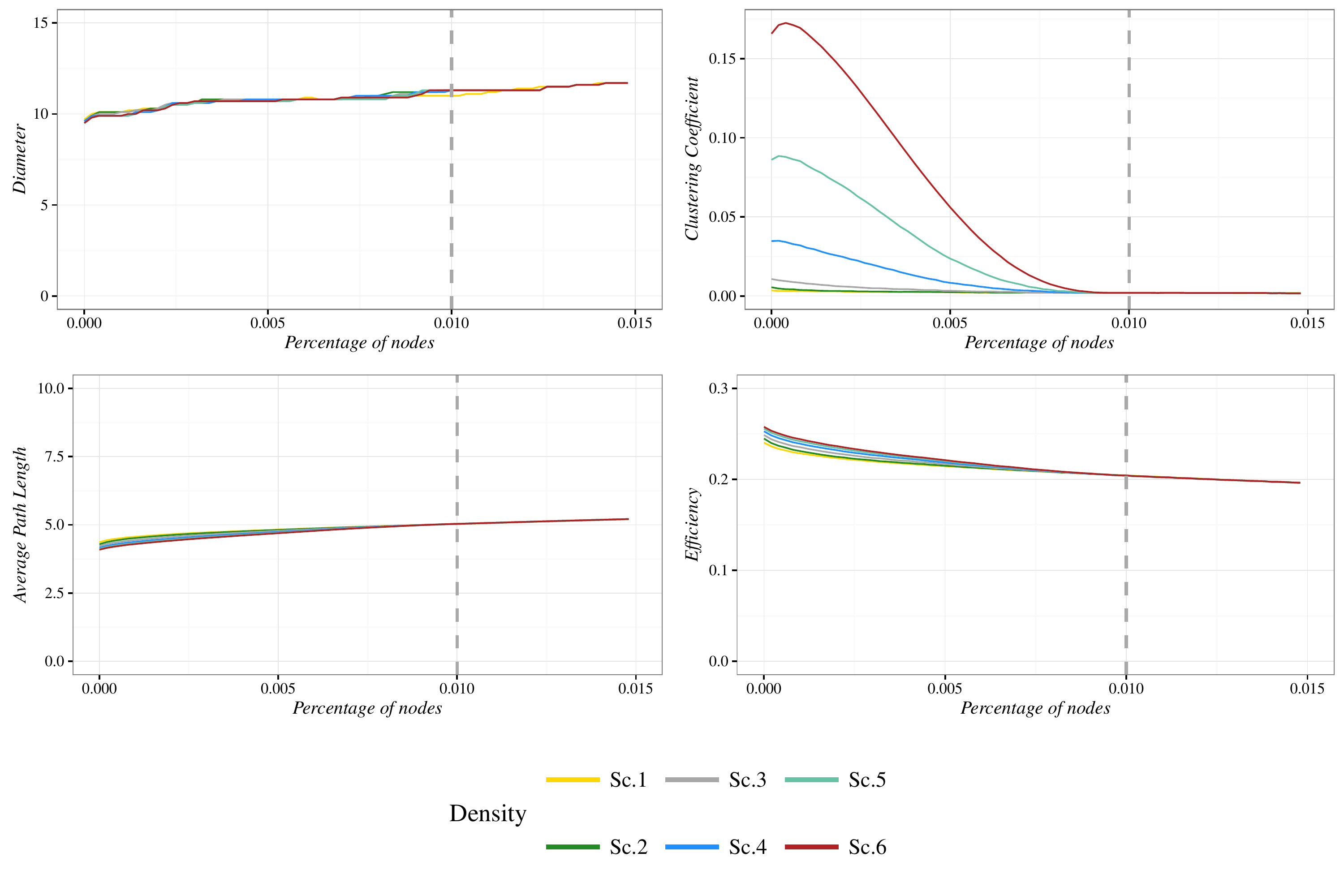}
%trim=3.5cm 24.3cm 3cm 3cm, clip = true
\caption{Resilience for simultaneous attack simulations with progressive manipulation of number of links in the network core; magnification of the area of Figure~\ref{attack} in which lays the rich-club.
The dashed line is placed in correspondence to the rich-club size.
All are results averaged over 10 instances.}
\label{attack_zoom}
\end{center}
\end{figure}

Analyzing Figures~\ref{attack} and~\ref{attack_zoom} we notice that the rich-club positively alters the initial statistics of the network and that its presence is not highly relevant with respect to simultaneous degree-targeted attack in networks that display a power-law degree distribution. In more detail, when we take into account scale-free networks, we observe that the overall trend of the considered measures is very close to that of the non-manipulated scale-free networks; in our case the curve with density $d= 0.09$ and related to Scenario 2. Indeed, the presence of the rich-club has an effect mainly on the initial values of the centrality measures and, in decreasing order of impact, on: the global clustering coefficient, the global efficiency, the average path length and the diameter.
The effect on all these metrics is amplified further by the density of the rich-club; thus, the higher its density the higher the overall centrality value. This is true in particular for the global clustering coefficient case in which, called $n_{rc}$ the number of nodes of the rich-club, are progressively generated up to $\binom{n_{rc}}{3}$ triangles, i.e. the number of triangles displayed by a complete subgraph of size $n_{rc}$. 

As previously mentioned, the effect of the rich-club is relatively strong for all the other centrality measures other than the diameter.
This is because the clustering coefficient, the efficiency and the average path length are measures averaged over all the network nodes (while the diameter is a more extremal measure), and are thus affected by the centrality values retained by the rich-club.
This bias is especially evident in scale-free networks whose heterogeneity in the degree distribution contributes to phenomena like the friendship paradox, which holds if the average degree of nodes in the network is smaller than the average degree of their neighbors \cite{eom2014generalized}.

The origin of the paradox is attributed to the existence of hub nodes and to the variance of the degree that contributes in altering the mean values of the degree over the neighborhoods of the nodes. Therefore, the observed deviations of the computed measures may be motivated by similar reasonings if we further consider the increase in the degree sequence variance induced by our manipulations. In summary, exacerbating the interconnections among hubs (i.e. to create progressively denser cores) has a relevant effect on the centrality measures averaged over the network nodes, but has no relevant effect in terms of resilience to a degree-targeted attack.
 
\begin{figure}[tbh]
\begin{center}
\includegraphics[scale = .5]{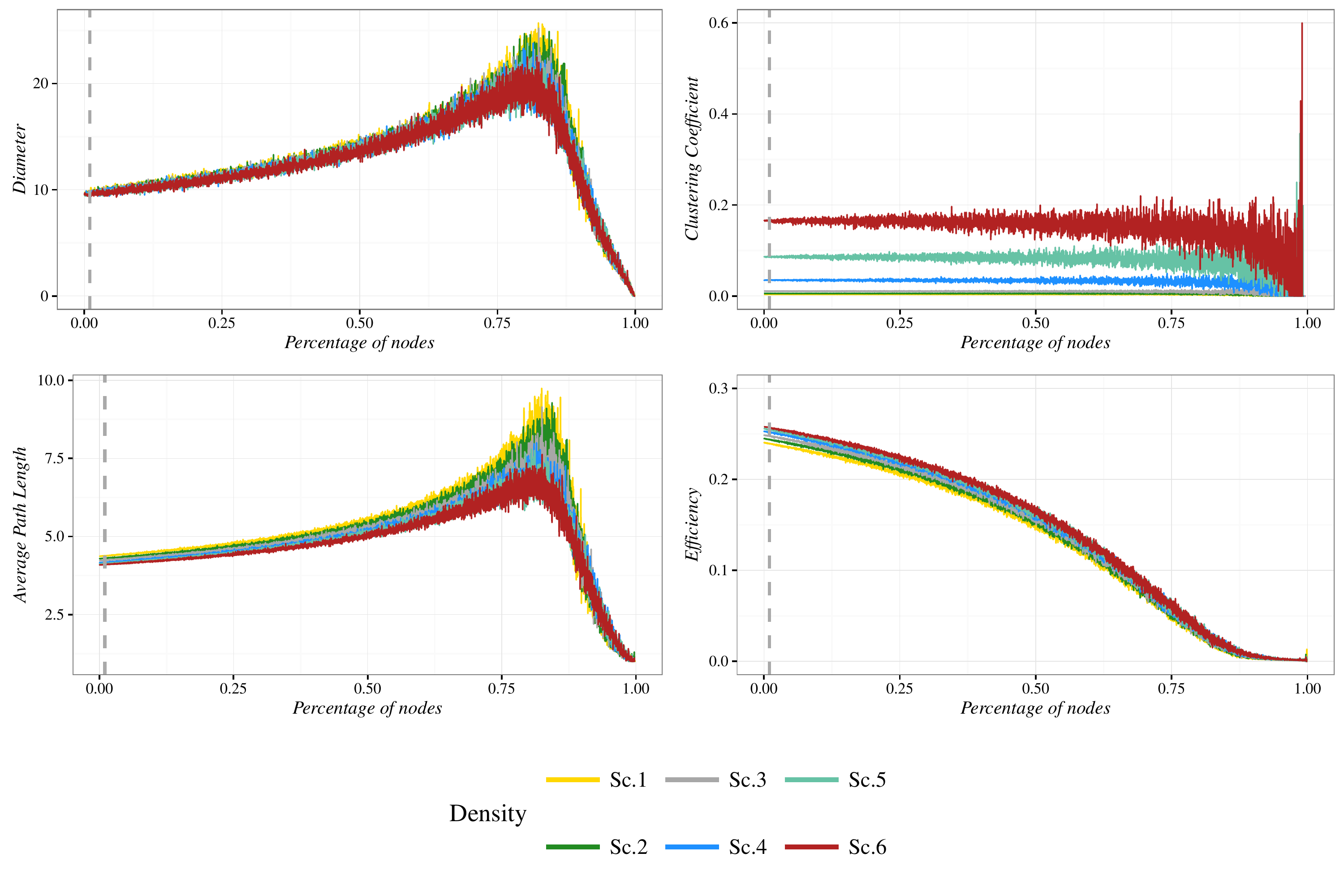}
%trim=3.5cm 24.3cm 3cm 3cm, clip = true
\caption{Resilience for simultaneous error simulations with progressive manipulation of the number of links in the network core.
The dashed line is placed in correspondence to the rich-club size.
All results are averaged over 10 instances.}
\label{error}
\end{center}
\end{figure}

In the case of error the rich-club in Figure~\ref{error}, according to its density, provides a very high fault tolerance to the considered system. Indeed, the nodes that  constitute the core make up a low portion ($1\%$) of the whole number of nodes and are thus less likely to be randomly removed. The low probability of hubs removal has an effect on the resilience of the system, which is guaranteed for all the observed measures. For instance, the diameter doubles only when about $75\%$ of the elements are removed, and the global clustering coefficient is kept during the simulations since the majority of triangles are located within the rich-club.

\begin{figure}[tbh]
\begin{center}
\includegraphics[scale = .5]{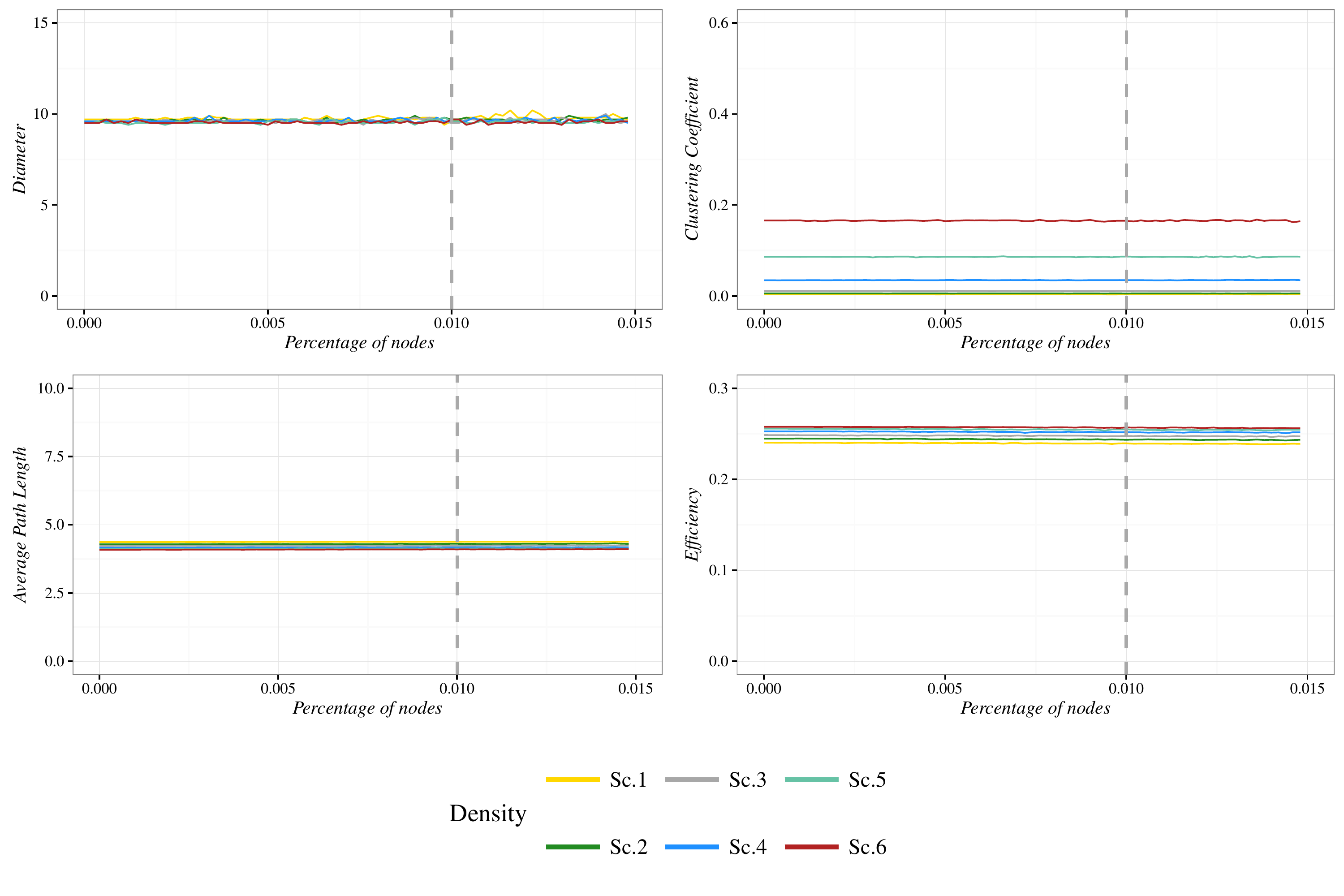}
%trim=3.5cm 24.3cm 3cm 3cm, clip = true
\caption{Resilience for simultaneous error simulations with progressive manipulation of the number of links in the network core; magnification of the area of Figure~\ref{error} in which lays the rich-club.
The dashed line is placed in correspondence to the rich-club size.
All results are averaged over 10 instances.}
\label{error_zoom}
\end{center}
\end{figure}

Figure~\ref{error_zoom} focuses on the area of the rich-club where the behavior of the considered measures follows a straight line, indicating a certain network stability for similar reasons as those discussed before.

\subsection{Periphery Thickening}
\label{P_T}

\begin{figure}[tbh]
\begin{center}
\includegraphics[scale = .5]{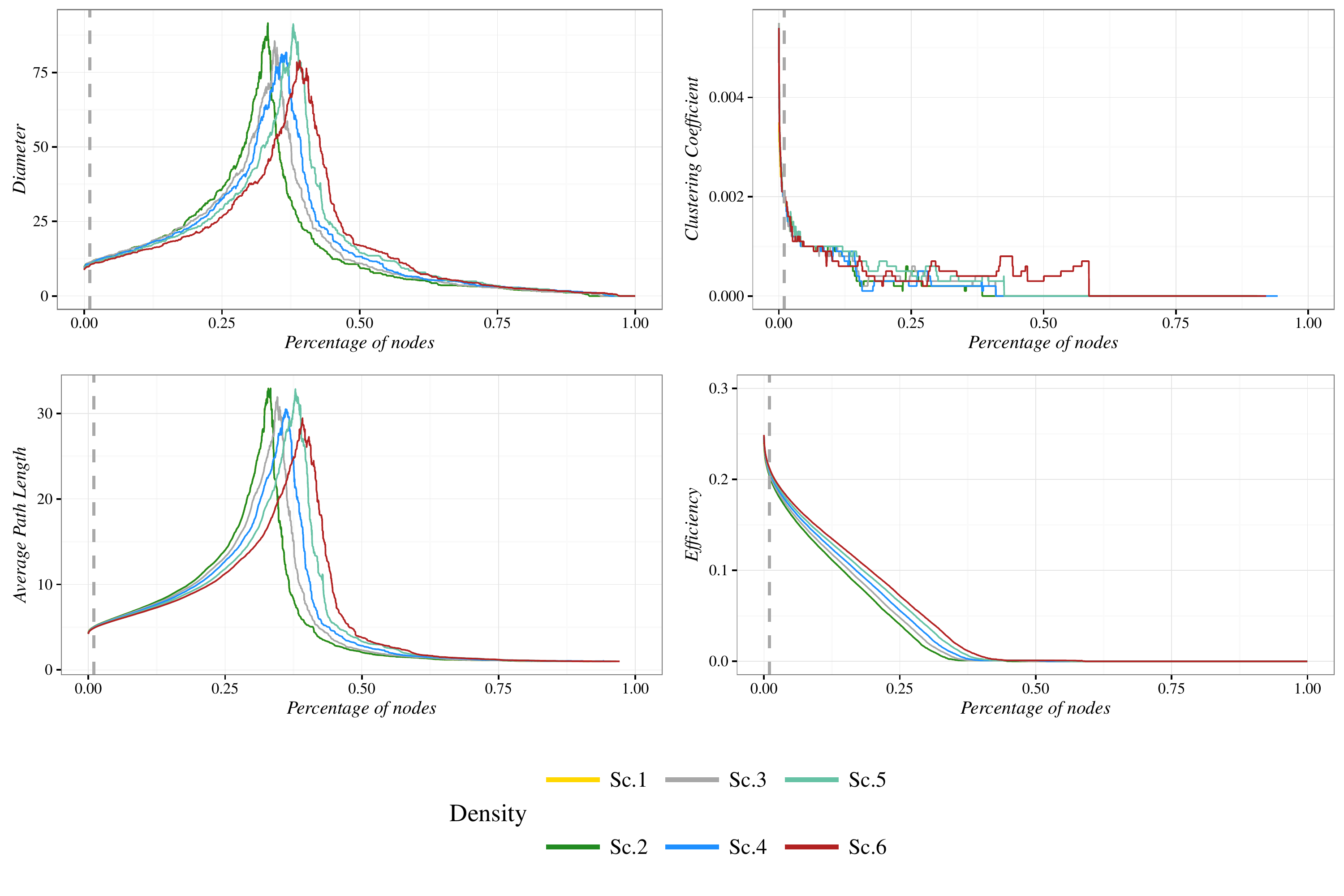}
%trim=3.5cm 24.3cm 3cm 3cm, clip = true
\caption{Resilience for simultaneous attack simulations with progressive manipulation of the number of links in the network periphery.
The dashed line is placed in correspondence to the rich-club size.
All results are averaged over 10 instances.}
\label{attack_p}
\end{center}
\end{figure}

As shown in Figure~\ref{attack_p}, networks with a denser periphery are more resilient to targeted attacks than networks with a denser core. When we look at the diameter and at the average path length, the peaks related to the two metrics occur in correspondence to a higher percentage of removed nodes and, differently from the case of core thickening, the number of added links has a role in determining the robustness to targeted removal.
This observation is consistent with the fact that, by adding links to the network periphery, we decrease the degree sequence variance, meaning we somehow regularize the considered networks. The obtained results recall the resilience to simultaneous degree-targeted attack in case of degree homogeneous networks \cite{crucitti2004error}. Additionally, the global clustering coefficient is much lower as links are not placed in order to thicken a small subgraph (the rich-club), consequently the likelihood to close a connected triple (to create a new triangle) is lower. Even in the case of global efficiency we observe a proportionately more resilient behavior across the number of added links.

\begin{figure}[tbh]
\begin{center}
\includegraphics[scale = .5]{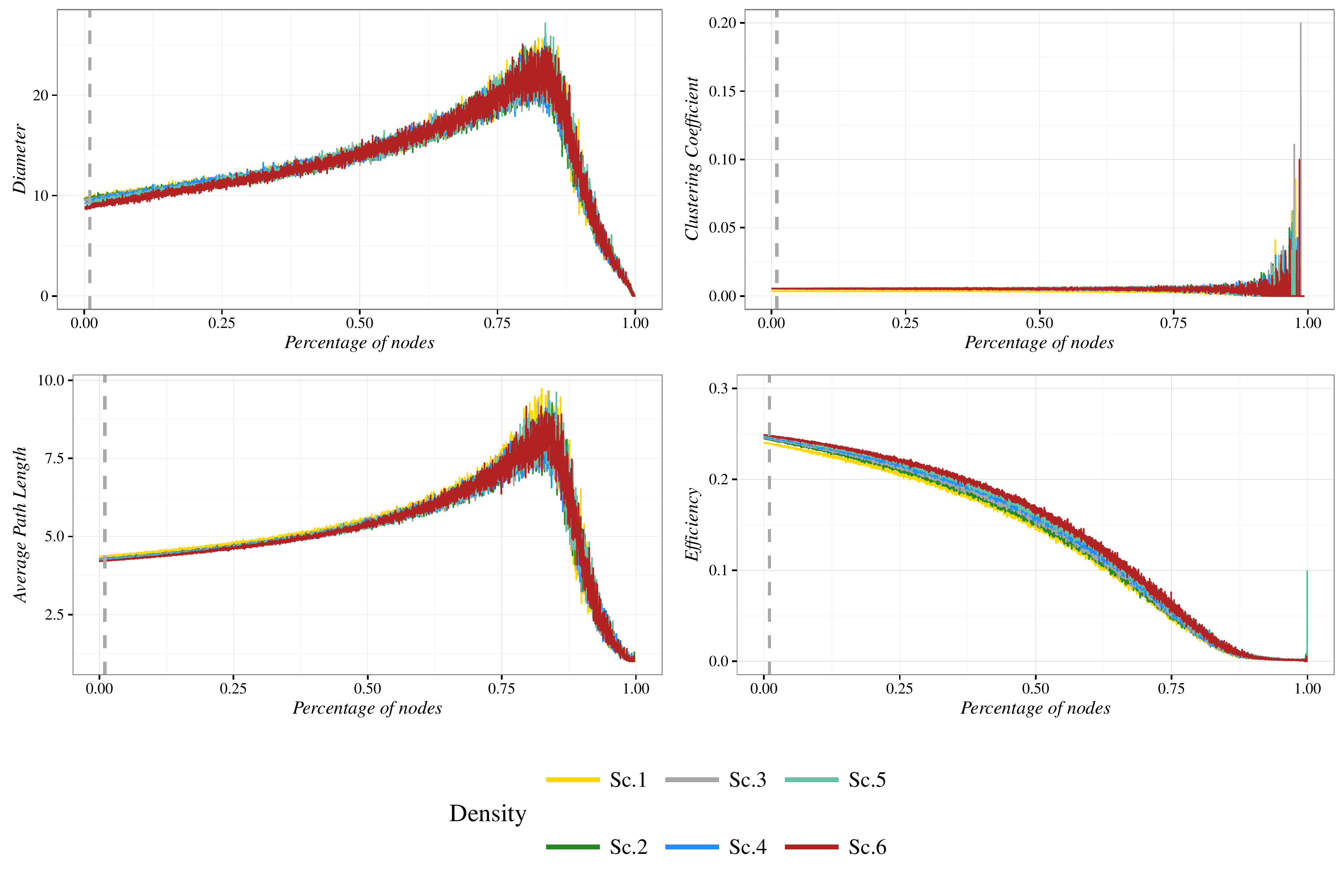}
%trim=3.5cm 24.3cm 3cm 3cm, clip = true
\caption{Resilience for simultaneous error simulations with progressive manipulation of the number of links in the network periphery.
The dashed line is placed in correspondence to the rich-club size.
All results are averaged over 10 instances.}
\label{error_p}
\end{center}
\end{figure}

In the case of error (see Figure~\ref{error_p}), the periphery thickening procedure leads to results that are similar to those of core thickening except for two considerations. The clustering coefficient is much lower, for the reasons discussed before, and the curves relating to different scenarios have similar and almost stacked trends; in other words, they refer to results that are comparable, despite the number of added links in the various scenarios being much different. This is because, as we lower the variance of the degree, the contribution of each node to the considered network metrics tends to be progressively the same. 

\section{Discussion and Conclusions}
\label{conclusion}

Herein we discuss the results of the simulations by looking at both their theoretical and practical meaning and implications. Considering two different perspectives regarding the results is helpful in better understanding the role of the rich-club in terms of network resilience and in providing insights into the demanding task of network supervision and engineering.

If we consider attack tolerance, the rich-club thickening initially guarantees a greater global cohesion predominantly in the core, as well as an overall better performance when removing the number of nodes below the $1\%$ threshold. Thus, the network provides better performance when only a few high-degree nodes are removed. The main drawback is that this high proportion of cohesion measure is retained by the nodes that are actually the most likely to be removed in the case of an attack. 

Considering attack tolerance once again, the periphery thickening has the main advantage in that it alters the network into a more resilient structure, which is able to keep its properties in the long run. This means that the network tends to maintain stable values of the performance measures when a high portion of the nodes is removed, since in this case the paths tend to be preserved. These aspects of network resilience are mainly regulated by the manipulation of the network degree-related heterogeneity (i.e by the manipulation of the variance of node degrees) that we perform through the procedures of core and periphery thickening.

In the case of error, the networks that display a dense core provide overall better performances that improve accordingly with the core density. Indeed, the trend of the different curves in the cases of core and periphery thickening are similar, but the former case provides also a better initial global efficiency and a higher value of clustering that lasts throughout the simulations.

Thus, if by looking at the simulations, a decision maker would evaluate where to put a set amount of links with respect to random node failure, the logical conclusion would be that it is better to increase the density of the network core and to increase that density as much as possible (compatibly with the amount of available links). 
The observations in cases of attack should be of different nature and should be weighted on an eventual foresight about the magnitude of possible attacks to the network. Indeed, if massive attacks on the network are possible, the periphery thickening (i.e. a network homogenization) should be preferred while if there is a higher likelihood of few hubs being removed, the core thickening (i.e. a network heterogenization), should be preferred. 

In other words, considering for instance the diameter, that is an extremal measure of communication, in the case of periphery thickening the curves have both shifted peaks and a lower slope according to the network density. It means that the network performance degenerates after a greater number of removed nodes and the considered performance measures are directly proportional to the network density. Indeed, for a fixed percentage of removed nodes the diameter is smaller as the density grows.

The concepts of attack magnitude and attack likelihood constitute two important aspects, related to risk profile of the network under observation that should be considered when different strategies of link addition are taken into account. 

However, these conclusions could be further discussed especially in case of resilience to massive attacks provided by networks treated with the periphery thickening procedure. Indeed, in case where about the $25\%$ of the network nodes (or more) are lost, issues regarding the performance could be discarded in favor of other issues regarding network recovery and catastrophes management. Thus, a decision maker may be not that interested in the performance measures from above once that the system has been dramatically disrupted. Using this consideration as a baseline, we may argue that once the percentage of removed nodes has passed such a right-shifted threshold, an advantage in terms of resilience is not particularly realistic due to the fact that any benefit can be only obtained once a loss of significantly large dimensions occurs. This may lead us to conclude that the core thickening procedure, i.e. the increase of the rich-club density, has to be considered as a practically better procedure to follow in order to enhance the network resilience.

In summary, the simulations highlight the relationship between the rich-club size and the attack magnitude, indicating that if the former is greater than the latter then a reasonable policy would be to perform a core thickening strategy.

Two aspects have to be considered further: on the one hand, the core thickening strategy provides a better resilience to errors and to small attacks (to hubs) but on the other this procedure, in accordance with the size and the density of the rich-club, exacerbates the degree-related asymmetry and thus entails a problem of equity of nodes that is invariably of interest in a number of real networks.
When the attack magnitude exceeds the rich-club size then simulations suggest a strategy of periphery thickening.

Therefore, a decision maker has to face controversial decisions regarding the adoption of a strategy that is affected by two parameters, the rich-club size and the attack magnitude, which are two measures generally difficult to obtain and foresee. 
%This sheds light on the importance of a better knowledge of the network structure and especially on the size of the rich-club of the network under observation together with other considerations related to the risk profile, i.e. the likelihood of an attack, on the considered network.
This reinforces the notion that a better understanding of the network structure and of the rich-club is relevant, especially when coupled with other concepts related to the risk profile and to the type of  system that is taken into account.

% DISCUSSIONE DI ALCUNI SVILUPPI FUTURI RISPETTO ALLA EQUITA' DEI NODI SOPRATUTTO IN NETWORK SOCIALI
%- DEL CASO WEIGHTED E DEL CASO DI LINK REMOVAL 
%- DEL CASO DI CASCADE FAILURES (FORSE PIU' INTERESSANTE) PERCHE' IN QUESTO CASO LA PRESENZA DEL RICH-CLUB POTREBBE PORRE DEI PROBLEMI DI PROPAGAZIONE 

%%%%%%%%%%%%%%%%%%%%%%%%%%%%%%%%%%%

%\section*{Acknowledgments} 

%\begin {thebibliography}{1}

%\bibliography{Alexandria_Cin}
%\bibliographystyle{plain}

%\end {thebibliography}

\end{document}